\documentclass[a4paper,fleqn,usenatbib]{mnras}

\usepackage[T1]{fontenc}
\usepackage{ae,aecompl}
\usepackage{comment}

\usepackage{graphicx}	
\usepackage{amsmath}	
\usepackage{amssymb}	
\usepackage{subfigure}
\usepackage{xfrac}
\usepackage{xcolor}

\title[Massive spheroids can form in single minor mergers]{Massive spheroids can form in single minor mergers}

\author[R. A. Jackson et al.]
{R. A. Jackson$^{1}$\thanks{E-mail: r.jackson9@herts.ac.uk},
G. Martin,$^{1,2}$
S. Kaviraj,$^{1}$
C. Laigle,$^{2}$
J. E. G. Devriendt,$^{2}$ \newauthor 
Y. Dubois, $^{3}$ and
C. Pichon $^{3,4,5}$
\\
$^{1}$Centre for Astrophysics Research, School of Physics, Astronomy and Mathematics, University of Hertfordshire, Hatfield, AL10 9AB, UK\\
$^{2}$Dept of Physics, University of Oxford, Keble Road, Oxford OX1 3RH UK\\
$^{3}$Institut d'Astrophysique de Paris, Sorbonne Universit\'es, UMPC Univ Paris 06 et CNRS, UMP 7095, 98 bis bd Arago, 75014 Paris, France\\
$^4$School of Physics, Korea Institute for Advanced Study (KIAS), 85 Hoegiro, Dongdaemun-gu, Seoul, 02455, Republic of Korea\\
$^5$Institute for Astronomy, University of Edinburgh, Royal Observatory, Blackford Hill, Edinburgh, EH9 3HJ, United Kingdom\\
}

\date{Accepted for publication in MNRAS}

\pubyear{2019}

\begin{document}
\label{firstpage}
\pagerange{\pageref{firstpage}--\pageref{lastpage}}
\maketitle

\begin{abstract}
Understanding how rotationally-supported discs transform into dispersion-dominated spheroids is central to our comprehension of galaxy evolution. Morphological transformation is largely merger-driven. While major mergers can efficiently create spheroids, recent work has highlighted the significant role of other processes, like minor mergers, in driving morphological change. Given their rich merger histories, spheroids typically exhibit large fractions of `ex-situ' stellar mass, i.e. mass that is accreted, via mergers, from external objects. This is particularly true for the most massive galaxies, whose stellar masses typically cannot be attained without a large number of mergers. Here, we explore an unusual population of massive ($M_*>$10$^{11}$M$_\odot$) spheroids, in the Horizon-AGN simulation, which exhibit anomalously low ex-situ mass fractions, indicating that they form without recourse to significant merging. These systems form in a single minor-merger event (with typical merger mass ratios of 0.11 - 0.33), with a specific orbital configuration, where the satellite orbit is virtually co-planar with the disc of the massive galaxy. The merger triggers a catastrophic change in morphology, over only a few hundred Myrs, coupled with strong in-situ star formation. While this channel produces a minority ($\sim$5 per cent) of such galaxies, our study demonstrates that the formation of at least some of the \textit{most massive} spheroids need not involve major mergers -- or any significant merging at all -- contrary to what is classically believed.
\end{abstract}




\section{Introduction}
Understanding the processes that alter the morphological mix of the Universe and, in particular, drive the formation of spheroidal galaxies, is a key topic in galaxy evolution studies. Observations indicate that, while the massive galaxy population in the early Universe was dominated by objects with discy morphologies \citep{Wuyts2011,Conselice2014,Buitrago2014,Shibuya2015}, their counterparts in the local Universe are dominated by systems that are spheroidal in nature \citep{Bernardi2003,Kaviraj2014a}. 

Morphological transformation, i.e. the conversion of rotationally-supported discs into slowly-rotating, dispersion-dominated spheroids, is believed to be triggered primarily by galaxy mergers \citep{Toomre1977,Negroponte1983,Barnes1992,DiMatteo2007,2009MNRAS.394.1956C,Oser2010,Dubois2016,Naab2014,RodriguezGomez2017,Welker2018,Martin2018a}. The past theoretical literature has often highlighted the important role of `major' mergers (mass ratios greater than $\sim 1:3$), as potentially the primary driver of morphological transformation \citep{Toomre1977,Negroponte1983,Solanes2015,Willett2015,Solanes2018}. The strong gravitational torques generated by nearly equal mass-ratio mergers efficiently move material from rotational orbits into the random, chaotic ones that make up spheroidal systems. This appears to be generally true, except in cases where the progenitors are extremely gas-rich, in which case residual gas from the merger can rebuild a disc, resulting in a remnant that still shows significant rotation \citep[e.g.][]{Springel2005,Font2017,Martin2018b}. 

Observations lend general support to this picture. The evolution of the morphological mix of galaxies shows a gradual increase in the fraction of spheroids over cosmic time \citep[e.g.][]{Butcher1984,Dressler1997,Conselice2014}. Many spheroids show signatures of rapid change in their stellar populations \citep[e.g.][]{Blake2004,Bundy2005,Ferreras2009,Wong2012,Carpineti2012,Kaviraj2014a,Wild2016}, and internal structure \citep[e.g.][]{Tacconi2008,McIntosh2008,Kaviraj2012a,Huertas2015,Rodrigues2017}, indicating a major merger in their recent history. 

However, {\color{black}both theoretical and observational work in the recent literature} \citep[e.g.][]{Zolotov2015,2015MNRAS.446.1957F,2017MNRAS.465.1241W,Martin2017,Martin2018b} has highlighted the potentially important role of minor mergers (mass ratios less than $\sim 1:3$), which, by virtue of their higher frequency, may significantly influence the evolution of massive galaxies. For example, many massive spheroids that are in the process of forming at $z\sim2$, the epoch at which the morphological mix of the Universe appears to change most rapidly \citep[e.g.][]{Conselice2014}, do not show tidal features and extended halos that should be visible if they were forming via major mergers \citep[e.g.][]{Williams2014,Lofthouse2017}. Indeed, recent studies have shown that (multiple) minor mergers are capable of re-distributing stellar orbits in galaxies and creating spheroidal, slowly rotating remnants \citep[e.g.][]{Bournaud2007,Qu2011,Taranu2013,Taranu2015,Moody2014}.

{\color{black}Minor mergers are capable of enhancing both star formation \citep[e.g.][]{Knapen2004,Ellison2008,Carpineti2012,Kaviraj2014a} and black-hole accretion rates \citep[e.g.][]{Kaviraj2014b,Pace2014} and, at low redshift, may account for more than half of the star formation budget in the Universe \citep{Kaviraj2014b}. Furthermore, the observed evolution in the sizes of massive (early-type) galaxies is also largely attributed to this process \citep[e.g.][]{Trujillo2006,Newman2012}. A growing body of evidence therefore points towards minor merging being a significant driver of not only morphological transformation but also of the stellar mass buildup, black-hole growth and size evolution of massive galaxies.}

Despite these examples, the impact of processes like minor mergers remains difficult to study observationally. This is because the surface-brightness of merger-driven tidal features scales with the mass ratio of the merger \citep[e.g.][]{Peirani2010,Kaviraj2010}, and thus the efficient identification of minor mergers typically requires deeper imaging than that offered by current large-area surveys \citep[e.g.][]{VD2005,Duc2011,Kaviraj2014b,Kaviraj2019}. Given this lack of survey depth, the effect of minor mergers on galaxy evolution remains relatively unexplored, although forthcoming instruments like LSST, EUCLID and JWST \citep[e.g.][]{2002SPIE.4836...10T,2017arXiv170801617R,2006SSRv..123..485G,2011arXiv1110.3193L} will enable the first truly statistical studies of minor-merger-driven galaxy evolution. 

Much of our current understanding of the role of this process {\color{black}therefore continues to come} from theoretical work e.g. from semi-analytical models \citep[e.g.][]{Somerville1999,Cole2000,Benson2003,Bower2006,Croton2006} and, more recently, from cosmological  hydro-dynamical simulations \citep[e.g.][]{Dubois2014,Vogelsberger2014,Schaye2015,Khandai2015,TaylorKobayashi2016,Kaviraj2017}. 

It is worth noting that exploring the role of various processes in driving morphological transformation ideally requires a simulation with a cosmological volume. While the past literature has often employed isolated and idealised simulations of galaxy mergers \citep[e.g.][]{Barnes1988,Hernquist1992,Bois2011}, such simulations exclude the broader effects of environment and gas accretion from the cosmic web. Furthermore, since the parameter space (e.g. orbital configurations, mass ratios etc.) explored by these studies is typically small, statistical conclusions about the importance of various processes that drive morphological change are difficult to draw. This can be particularly important for channels, like the one studied here, that are relatively rare.

The consensus view from simulations is that morphological transformation and the creation of spheroids can proceed via various channels and shows a complicated dependence on factors such as mass ratios, gas fractions, environments and orbital configurations (see the study by \citet{Martin2018a} which quantifies these dependencies in detail). Here, we describe a channel for spheroid formation in which a \textit{single} minor merger creates a massive ($M_*>$ 10$^{11}$M$_\odot$) spheroidal system. In this process, the satellites approach the more massive merging progenitor on a trajectory that is close to being co-planar to the plane of the massive disc. Notwithstanding the low mass ratios of many of these mergers, these events are able to trigger a catastrophic change in morphology over just a few hundred Myrs, coupled with an episode of strong in-situ star formation, that creates a slowly-rotating, massive spheroidal system from an otherwise typical disc galaxy. 


This paper is organised as follows. In Section \ref{horizon}, we briefly describe the Horizon-AGN simulation and how galaxies, mergers and morphologies are defined. In Section \ref{sample}, we describe the properties of the spheroidal galaxies studied here and follow the evolution of their stellar and gas mass and morphology over cosmic time. In Section \ref{progenitor}, we explore the properties of the progenitor systems (e.g. the gas fraction and clumpiness of the more massive galaxy and the orbital configuration of the merger) and investigate how such spheroids form via single minor-merger events. We summarise our findings in Section \ref{conc}. 

\section{Horizon-AGN Simulation} \label{horizon}
In this study, we use the Horizon-AGN cosmological hydrodynamical simulation \citep{Dubois2014}, which employs \textsc{ramses} \citep{2002A&A...385..337T}, an adaptive mesh refinement (AMR) hydrodynamics code. Horizon-AGN simulates a 100 $h^{-1}$coMpc$^3$ volume, with WMAP7 $\Lambda$CDM initial conditions \citep{2011ApJS..192...18K}. The simulation contains $1024^3$ dark matter particles on an initial 1024$^3$ cell gas grid. This is refined according to a quasi Lagrangian criterion, when 8 times the initial total matter resolution is reached in a cell. This refinement can continue until 1 kpc in proper units. Horizon-AGN has  dark-matter and stellar-mass resolutions of $8\times 10^7$ M$_{\odot}$ and $4\times 10^6$ M$_{\odot}$ respectively. 

The simulation includes prescriptions for feedback from both stars and AGN. Continuous stellar feedback is employed including momentum, mechanical energy and metals from stellar winds, Type Ia and Type II supernovae (SNe). Type Ia SNe are implemented following \citet{1986A&A...154..279M}, assuming a binary fraction of 5\% \citep{2001ApJ...558..351M}. Feedback from Type II SNe and stellar winds is implemented using \textsc{starburst99} \citep{1999ApJS..123....3L,2010ApJS..189..309L}, which employs the Padova model \citep{2000A&AS..141..371G} with thermally pulsating asymptotic branch stars \citep{1993ApJ...413..641V}. 

Black Holes (BHs) are implemented as `sink' particles, with initial masses of $10^5$M$_\odot$, and grow via gas accretion or coalescence with other BHs. When gas accretes onto a central BH it imparts feedback on ambient gas via two methods, depending on the accretion rate. When there is a high ratio of gas accretion to the Eddington rate (Eddington ratios of $> 0.01$), 1.5 per cent of the energy is injected into the gas as thermal energy (a `quasar' mode). However, when Eddington ratios are less than 0.01, bipolar jets are employed with velocities of $10^4$ km s$^{-1}$, which constitutes a `radio' mode with an efficiency of 10 per cent. For more details of the baryonic recipes used in the simulation, we refer readers to \citet{Dubois2014,Dubois2016} and \citet{Kaviraj2017}. 

{\color{black}Horizon-AGN produces good agreement with an array of observational quantities that trace the aggregate growth of stars and BHs, particularly in massive galaxies. For example, in the stellar mass range $M_*>$10$^{10.5}$M$_\odot$, it reproduces the morphological mix of the nearby Universe \citep{Dubois2016,Martin2018b}, the evolution of the stellar mass/luminosity functions, rest-frame UV-optical-near-infrared colours, the cosmic star formation history and the star formation main sequence \citep{Kaviraj2017,Martin2017}, galaxy merger rates at $z>1$ \citep{Kaviraj2015} and the evolving demographics of BHs over cosmic time \citep{Volonteri2016}. However, the simulation overproduces the number densities of very low-mass galaxies at low and intermediate redshift (which is typical of all simulations in this class of models). The rest-frame UV colours of galaxies in the red sequence are also typically too blue, indicating that feedback processes employed in the model do not suppress star formation completely in `passive' galaxies. Note, however, that since the star formation rate density in the red sequence is negligible compared to that in the blue cloud this does not affect the overall reproduction of galaxy properties.}

\subsection{Identifying galaxies and mergers} 
{\color{black}In each snapshot of the simulation galaxy catalogues are produced using the \textsc{adaptahop} structure finder \citep{Aubert2004} applied to the star particles}. Structures are selected when the local density, calculated for each particle using the nearest 20 neighbours, exceeds the average matter density by a factor of 178. A minimum of 50 particles is required for a structure to be identified, which imposes a minimum galaxy stellar mass of $\sim2\times10^{8}$M$_{\odot}$. {\color{black}Merger trees are produced for each galaxy from $z=0.06$ to $z=3$, with a typical spacing between time steps of $\sim130$ Myr, using the method described in \citet{Tweed2009}.}

\subsection{Morphologies}
Following \citet{Martin2018a}, we define galaxy morphology using stellar kinematics, specifically $\sfrac{V}{\sigma}$, which is the ratio between the mean rotational velocity ($V$) and the mean velocity dispersion ($\sigma$). Galaxies with higher values of $\sfrac{V}{\sigma}$ are rotationally-supported i.e. disc galaxies, while those with lower $\sfrac{V}{\sigma}$ values are pressure-supported spheroidal systems.  

In order to obtain $\sfrac{V}{\sigma}$, the coordinate system is rotated so that the $z$-axis is oriented along the stellar angular-momentum vector. $V$ is then defined as $V_{\theta}$, the mean tangential velocity component in cylindrical co-ordinates. The velocity dispersion ($\sigma$) is calculated by taking the standard deviations of the radial ($\sigma_{r}$), tangential ($\sigma_{\theta}$) and vertical star particle velocities ($\sigma_{z}$) and summing them in quadrature. $\sfrac{V}{\sigma}$ is then defined as:

\begin{equation}
\frac{V}{\sigma} = \frac{\sqrt{3} \bar{V_{\theta}}}{\sqrt{\sigma^2_r + \sigma^2_z + \sigma^2_\theta}}
\end{equation}

{\color{black}As described in \citet{Martin2018a}, galaxies are separated morphologically by considering a range of values for $\sfrac{V}{\sigma}$ and comparing how the corresponding predicted early-type fractions compare to local observations \citep{Conselice2006}. The value of $\sfrac{V}{\sigma}$ which produces the best agreement with the observed morphological mix of the Universe is then used to separate early- and late-type galaxies. This $\sfrac{V}{\sigma}$ threshold is 0.55 \citep[see Figure 1 in ][]{Martin2018a} i.e. galaxies with $\sfrac{V}{\sigma}>0.55$ are considered to be discs, while those with $\sfrac{V}{\sigma}<0.55$ are spheroids. Note that the systems we study in this paper have $\sfrac{V}{\sigma}$ values that are significantly lower than 0.55 i.e. these galaxies are firmly in the spheroid regime. We note that the highest $\sfrac{V}{\sigma}$ in Horizon-AGN is 2.11, which is lower than observed values. This is because the spatial resolution results in the bulge components of discs being slightly too massive.}


\section{Spheroidal galaxies with low ex-situ mass fractions} 
\label{sample}

\begin{figure}
\includegraphics[width=0.5\textwidth]{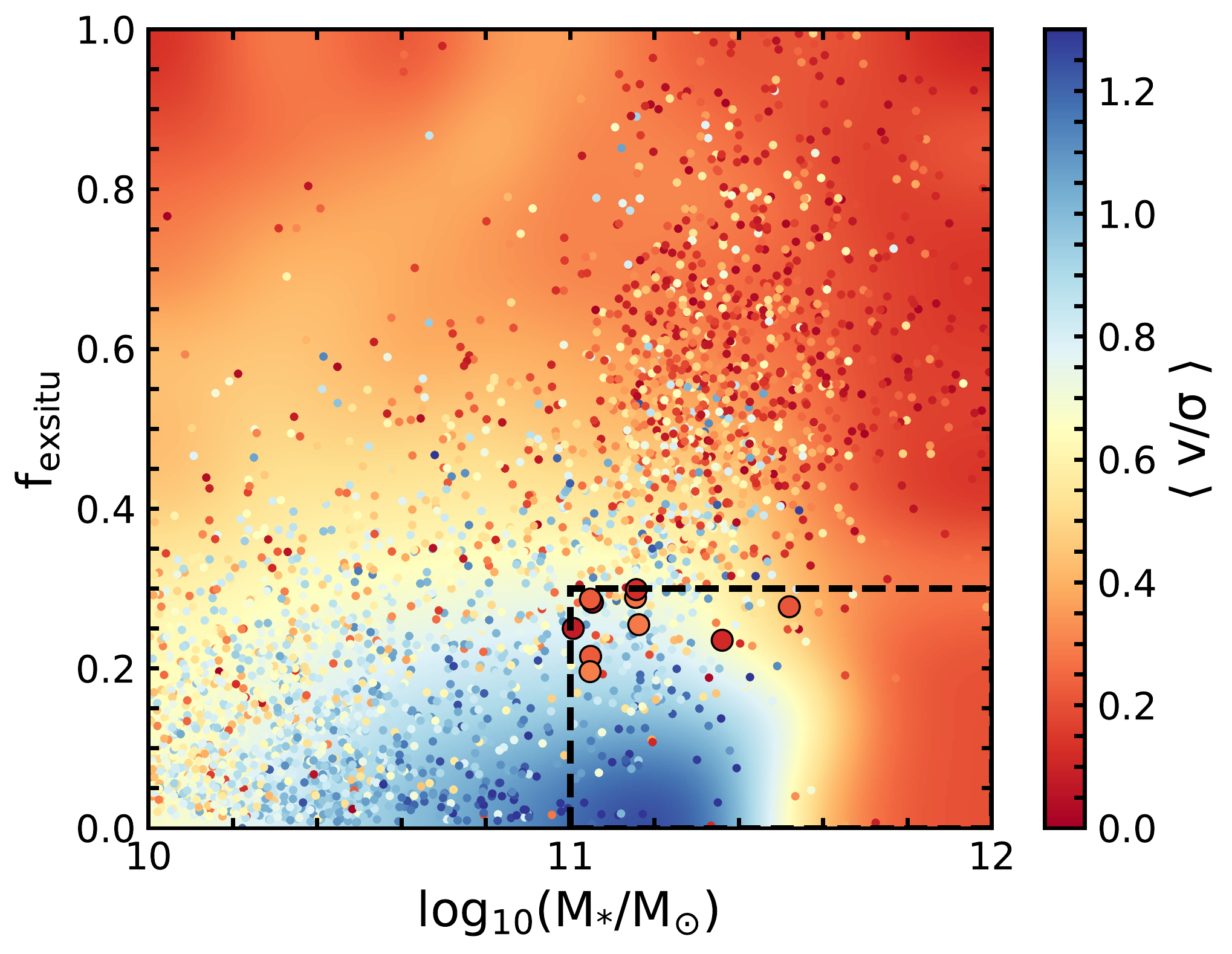}
\caption{2D histogram showing the fraction of ex-situ mass vs stellar mass of galaxies in the Horizon-AGN simulation at $z=0.06$. The individual points show a random selection of galaxies and the background colour is derived using a cubic spline interpolation over all galaxies. The colour coding indicates the value of $\sfrac{V}{\sigma}$ of both the background and the individual points. Massive spheroids typically exhibit large fractions of ex-situ stellar mass. The spheroids in this study have $\sfrac{V}{\sigma}$ $< 0.3$, $f_{exsitu}$ $< 0.3$ and stellar mass $>$ $10^{11}$M$_\odot$, and are shown using the large outlined points in the bottom right-hand corner of the plot. Note that we do not consider objects that may have $\sfrac{V}{\sigma}$ $< 0.3$ but are in ongoing mergers, as the $\sfrac{V}{\sigma}$ values of such objects are highly transient and the morphology of the remnant is uncertain.
}
\label{selection}
\end{figure}

\subsection{Sample selection and properties of the galaxies}
\label{properties}
In a recent study \citet{Martin2018b} have used Horizon-AGN to perform a detailed study of how mergers drive the change in the morphological mix of the Universe over cosmic time. Not unexpectedly, spheroidal galaxies, particularly those more massive than the knee of the galaxy mass function (M$_*$$>$10$^{11}$M$_\odot$), typically have rich merger histories. As a result, they usually exhibit very high ex-situ mass fractions, where ex-situ mass is defined as that directly accreted from external objects, rather than having been formed in-situ in the main (i.e. most massive) branch of the merger tree of the galaxy in question \citep{Martin2018b}. Figure \ref{selection} shows a 2-D histogram of galaxies in Horizon-AGN at the last snapshot ($z=0.06$), with the ex-situ stellar mass fraction plotted against stellar mass. Points indicate individual galaxies, while colours indicate their $\sfrac{V}{\sigma}$ ratios.  

While most massive spheroids show high ex-situ mass fractions, there exists a small, `anomalous' population of spheroidal systems with unusually low ex-situ mass fractions ($<$0.3), that are more typical of discs. These objects are the focus of this study. We select galaxies that (1) have stellar masses greater than $10^{11}$M$_\odot$, which puts them beyond the knee of the galaxy mass function \citep[see e.g.][]{Kaviraj2017} where major mergers have classically been thought to be essential for building up galaxy mass \citep[e.g.][]{Bell2004,Bundy2006,Faber2007}, (2) exhibit $\sfrac{V}{\sigma}$ $<0.3$ i.e. are firmly in the spheroid regime and (3) have very low ex-situ mass fractions ($f_{exsitu}$ $<0.3$). The $f_{exsitu}$ threshold is not arbitrary but chosen at the approximate mass-ratio demarcation between major and minor mergers ($\sim 1:3$), making it unlikely that these galaxies have undergone major mergers. As we show below, while they have not undergone major interactions, these systems are the products of single minor mergers with a specific set of properties. {\color{black}Note that simulated galaxies in this mass regime have very complete merger histories. For example, the merger histories of galaxies with $M_*>$10$^{11}$M$_\odot$ are close to 90 per cent complete for mass ratios greater than $\sim1:10$ out to $z\sim3$, and almost 100 per cent complete out to $z\sim1$ \citep[see Figure 1 in][]{Martin2018b}.} 

\begin{figure}
\includegraphics[width=0.45\textwidth]{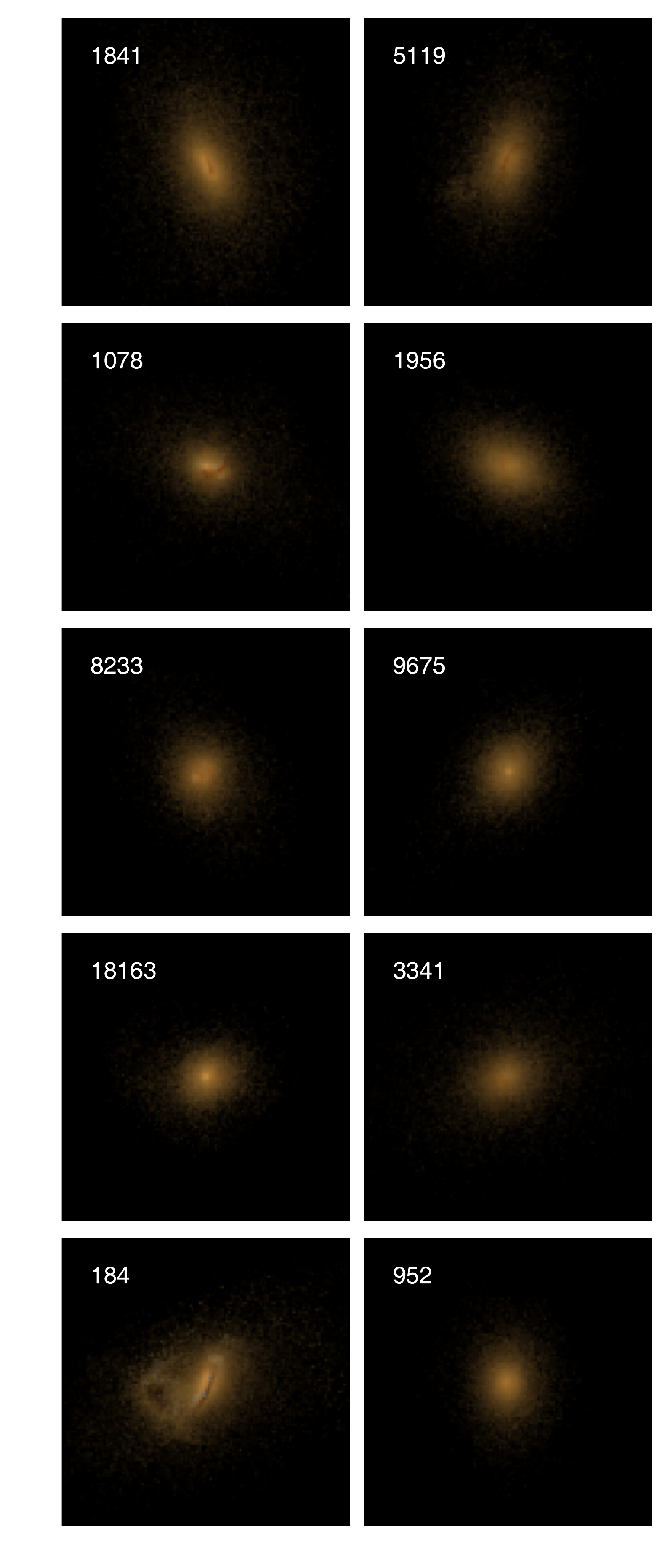}
\caption{Mock $gri$ images of the spheroids studied in this paper. The ID of each galaxy (same as in the tables) is shown in the panels. The procedure for constructing the images is described in Section 3.1. The angular resolution of the images is 1.05 arcsec/pixel.} 
\label{pics of gals}
\end{figure}

\begin{center}
\begin{table*}
\centering
\begin{tabular}{||c | c | c | c | c | c | c | c | c | c | c |||}
1 & 2 & 3 & 4 & 5 & 6 & 7 & 8 & 9 & 10 & 11\\
\hline
ID & M$_*$ & M$_*$/M$_i$ & $\sfrac{V}{\sigma}$ & R$_{e}$ & $\mu_e^r$ & Density & f$_{exsitu}$ & Merger $z$ & Merger mass & Rel. $\phi_{max}$\\
       & (10$^{11}$M$_\odot$) &  & & (kpc) & (mag arcsec$^{-2}$) & perc. & & & ratio &  \\
\hline
\hline
1841   & 2.28   & 1.97  & 0.12 & 10.39  & 21.08  & 0.30  & 0.24  & 0.4  & 0.15 & 63.7 \\
5119   & 1.42   & 1.51  & 0.28 & 12.28  & 22.43  & 0.26  & 0.29 & 0.95  & 0.11 & 20.52 \\
1078   & 1.33   & 6.33  & 0.10 & 10.87  & 21.55  & 0.82  & 0.29 & 2.45  & 0.33 & 27.22 \\
1956   & 1.45   & 3.63  & 0.27 & 12.43  & 22.22  & 0.75  & 0.25 & 1.3   & 0.17 & 37.02 \\
8233   & 1.11   & 1.54  & 0.24 & 7.66   & 21.57  & 0.53  & 0.21 & 0.3   & 0.24 & 18.51 \\
9675   & 1.01   & 1.77  & 0.07 & 9.26   & 21.69  & 0.10  & 0.25 & 0.6   & 0.17 & 9.75 \\
18163  & 1.11   & 1.71  & 0.30 & 7.34   & 20.60  & 0.17  & 0.20 & 0.25  & 0.20 & 8.16 \\
3341   & 1.13   & 3.55  & 0.03 & 11.53  & 22.83  & 0.77  & 0.28 & 2.1   & 0.15 & 9.12 \\
184    & 3.36   & 2.85  & 0.16 & 10.07  & 19.75  & 0.92  & 0.28 & 0.9   & 0.27 & 31.38 \\
952    & 1.13   & 2.69  & 0.23 & 8.98   & 21.95  & 0.04  & 0.29 & 0.4   & 0.20 & 32.82 
\end{tabular}
\caption{Properties of the spheroids studied in this paper (cols 1-8) and properties of the mergers that create them (cols 9-11). Columns are as follows: (1) Galaxy ID (2) Stellar mass in units of 10$^{11}$M$_\odot$ (3) Ratio of stellar mass of the spheroid to the mass of the progenitors before the merger (4) $\sfrac{V}{\sigma}$ (recall that the boundary between discs and spheroids is $\sfrac{V}{\sigma}\sim 0.55$) (5) Effective radius ($R_e$) in kpc (6) Effective surface brightness (i.e. the average $r$-band surface-brightness within $R_e$) (7) Local environment, expressed in terms of the percentile in local density (values greater than 0.9 indicate cluster environments, values in the range 0.4 - 0.9 indicate groups and values below 0.4 indicate the field) (8) The fraction of ex-situ stellar mass  (9) Redshift of the merger (10) Stellar mass ratio of the merger and (11) Ratio of the maximum star-formation rate in the merger remnant during the phase where the morphological transformation is fastest (shown by the grey regions in Figure \ref{1956merg}) and the average value of the observational star formation main sequence, for $M_*=$10$^{11}$M$_\odot$, at the redshift of the merger that creates these spheroids.}
\label{spheroid properties}
\end{table*}
\end{center}


Our initial selection yields 20 galaxies, half of which exhibit, on visual inspection of their images, strong morphological disturbances (i.e. are in ongoing mergers). As the $\sfrac{V}{\sigma}$ of merging objects can be extremely spurious \citep{Martin2018b}, the final values of $\sfrac{V}{\sigma}$ for these systems are uncertain. Thus, we study only the 10 systems (indicated by the large black points in Figure \ref{selection}) that are morphologically settled spheroids. This represents $\sim 5$ per cent of spheroids {\color{black}($\sfrac{V}{\sigma} < 0.55$)} that have $M_*>$10$^{11}$M$_\odot$. 

Figure \ref{pics of gals} shows mock $gri$ images of these galaxies (note the clearly spheroidal morphologies). To produce these images we use the \citet{Bruzual2003} stellar population synthesis models, with a \citet{Chabrier2003} initial mass function, to calculate intrinsic spectral energy distributions (SEDs) of the constituent star particles within each galaxy. Dust-attenuated SEDs are then calculated using the \texttt{SUNSET} code, as described in \citet{Kaviraj2017}, with line-of-sight optical depths calculated assuming a dust-to-metal ratio of 0.4 \citep[e.g.][]{Draine2007}. The attenuated SEDs of star particles are then convolved with the SDSS $g$, $r$ and $i$ band filter response curves to produce mock images of each galaxy. {\color{black}The images include a Gaussian point spread function with 1.2 arcsecond FWHM, but do not include a background or shot noise}.

This population of galaxies consists of extremely massive, slowly-rotating spheroids in the local Universe that have formed without recourse to a significant amount of merging, since only a small fraction of their stellar mass has been directly accreted from external sources. It is worth noting that the cuts in $\sfrac{V}{\sigma}$ and $f_{exsitu}$ made here are conservative i.e. we intentionally select very low values of $\sfrac{V}{\sigma}$ and $f_{exsitu}$ in order to cleanly isolate systems that have virtually no mergers in their assembly histories, yet are both massive and spheroidal. It could be reasonably expected, however, that the processes that drive the formation of these spheroids may also play some role in the creation of spheroids which have higher (i.e. more typical) values of $\sfrac{V}{\sigma}$ and $f_{exsitu}$. 

In Table \ref{spheroid properties}, we summarise key properties of these spheroids: the stellar mass today, the fractional increase in stellar mass after the merger that creates these systems, the $\sfrac{V}{\sigma}$, effective radius, effective surface-brightnesses in the $r$-band\footnote{The effective surface brightness is defined as the mean surface brightness within an effective radius.}, the local environment at the present day and the peak star formation rate (SFR) as a fraction of the average SFR of an equal mass galaxy at the redshift of the merger (`Rel. $\phi_{max}$'). 

Following \citet{Martin2018a}, we estimate local environment by ranking each galaxy by its local number density, which is calculated using an adaptive kernel density estimation method \citep{Breiman1977}. Galaxies are then sorted into percentiles of density, so that galaxies in the 0-40$^{\mathrm{th}}$ percentile range inhabit environments that roughly correspond to the `field', those in the 40-90$^{\mathrm{th}}$ percentile range inhabit `groups' and those in the 90-100$^{\mathrm{th}}$ percentile range inhabit environments typical of `clusters'. We refer readers to \citet{Martin2018a} for more details of this procedure. Rel. $\phi_{max}$ is calculated by dividing the maximum spheroid SFR by the average value of the observational star-formation main sequence for $M_*=$10$^{11}$M$_\odot$ at the redshift of the merger that creates these spheroids. {\color{black}The star-formation main sequences are taken from \citet{Lee2015} ($0.3 < z < 1.2$), \citet{Karim2011} ($0.3 < z < 2.5$), \citet{Whitaker2012} ($0.3 < z < 2.5$) and \citet{Tasca2015} ($2.5 < z < 5.0$).}

Table \ref{spheroid properties} shows that the current mass of the spheroids is several factors ($\times$1.5 - 6.3) larger than the mass of their progenitor systems. The effective radii ($R_e$) are typically larger than the typical value of $R_e$ for spheroids of such masses in the present-day Universe ($\sim$4 kpc; \citet{Bernardi2010}) and the effective surface-brightnesses of these objects are somewhat fainter than that of typical massive spheroids \citep[$\sim$18 mag arcsec$^{-2}$, e.g.][]{Blanton2005,Driver2005}. Indeed, the effective surface-brightness of some of these systems is close to the surface-brightness limit of current large surveys like the SDSS ($\sim$23 mag arcsec$^{-2}$, e.g. \citet{Kniazev2004}). The local environments of our spheroids span the full range of density percentiles, from systems that are in the field to one object which is in a cluster. These events are, therefore, not strongly correlated with a particular type of local environment. Finally, the SFRs of these spheroids show significant enhancements above the average SFR of similar-mass galaxies at the redshifts of their mergers.


\begin{figure*}
\centering
\includegraphics[width=0.9\textwidth]{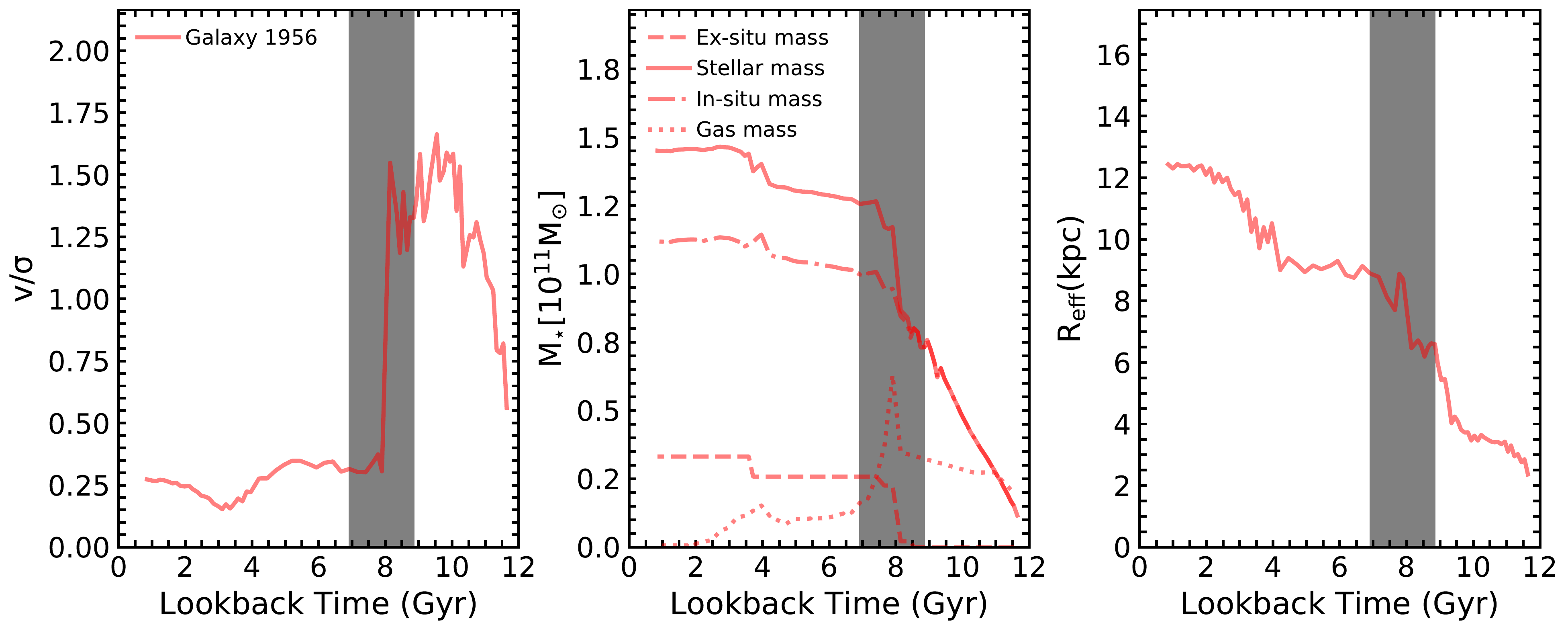}
\includegraphics[width=0.9\textwidth]{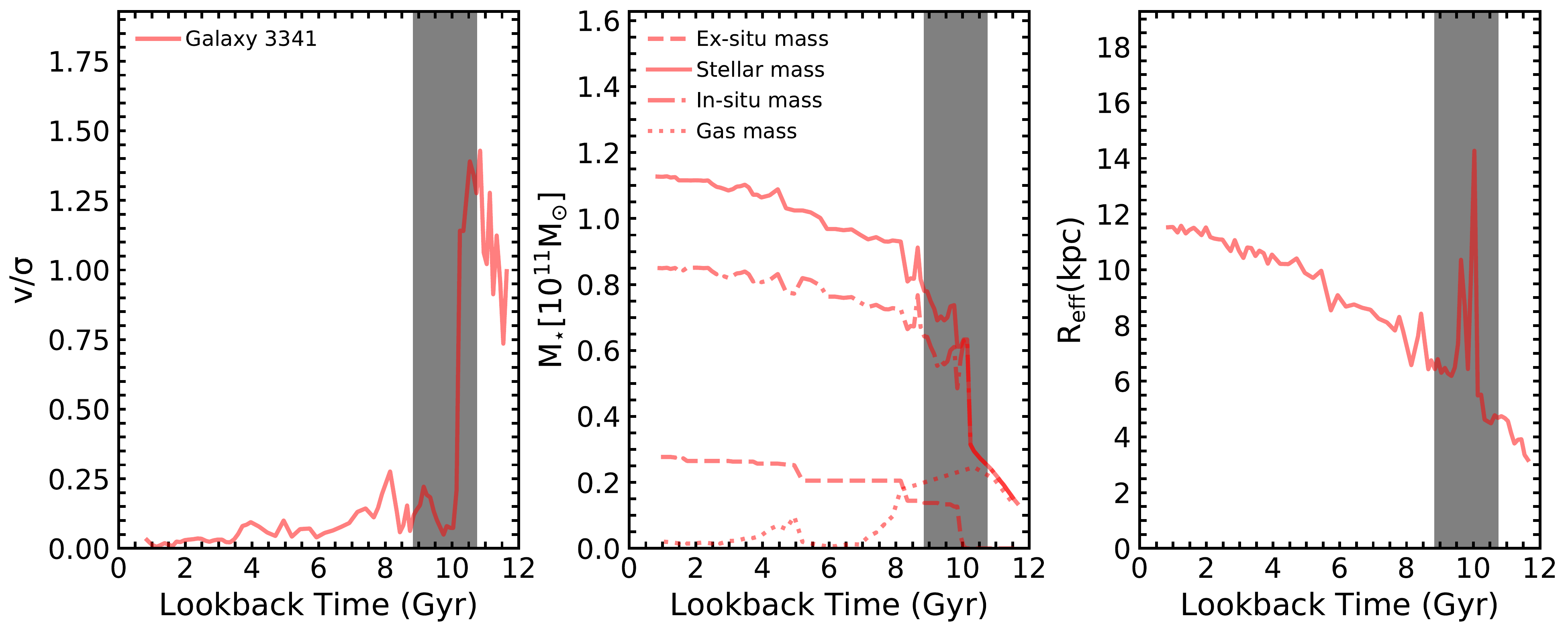}
\caption{The evolution of the properties of the progenitor system of the spheroids in this study. The top and bottom rows show the evolution of two typical galaxies in our sample. The grey region indicates the epoch at which the merger event takes place. Left: Evolution in $\sfrac{V}{\sigma}$ of the more massive galaxy. Middle: Change in the stellar mass (solid), ex-situ (i.e accreted) mass (dashed), in-situ mass (dashed-dotted) and gas mass (dotted) of the more massive galaxy. The near-step change in the ex-situ mass around the time of the merger indicates the stellar mass brought in by the accreted satellite. Right: Change in the effective radius of the more massive galaxy. In all cases, the merger results in a catastrophic (and permanent) decrease in $\sfrac{V}{\sigma}$, followed by a period of significant star formation and an increase in the effective radius of the galaxy.}
\label{1956merg}
\end{figure*}

\subsection{Creation in single minor mergers} 
\label{creation}


Table \ref{spheroid properties} also summarises the properties of the mergers that create the spheroids in this study. The redshifts of these events do not show a preference for a particular epoch. The mass ratios indicate that the events are minor mergers, albeit with a wide range of values, ranging from events that are close to being major mergers (mass ratios close to $\sim$1:3) to those where the mass ratios are as low as $\sim$1:10. {\color{black}It is worth noting that many of these spheroids exhibit prominent dust lanes (see Figure \ref{pics of gals}) which are considered to be signposts of recent minor mergers \citep[see e.g.][]{Kaviraj2012a}.}

We proceed by exploring the evolution of the properties of the more massive galaxy (stellar mass, ex-situ stellar mass, gas mass, effective radius and $\sfrac{V}{\sigma}$) in the minor mergers that create these spheroids. 
We follow these properties between $z=0.06$ and $z=3$ and focus on changes in these characteristics before and after the minor-merger event in question. Note that, in all cases, the massive galaxy in these mergers is firmly in the disc regime \textit{before} the merger takes place. 

In Figure \ref{1956merg} we show the evolution of these properties for two galaxies, in order to illustrate the process (the evolution is similar for other spheroids in our sample). In each panel, we highlight in grey the redshift at which the merger occurs. The left-hand panels show the evolution in $\sfrac{V}{\sigma}$. After the event the galaxies undergo a catastrophic decrease in $\sfrac{V}{\sigma}$ which can drop from the disc regime $\sfrac{V}{\sigma}>0.55$ well into the spheroid regime within only a few hundred Myrs (which represents only a few dynamical timescales). The $\sfrac{V}{\sigma}$ ratio does not recover after the merger, indicating that a disc does not reform after the event. We note that, while some of these spheroids do undergo further mergers (typically events with very low mass ratios), the creation of the dispersion-dominated system takes place in a \textit{single} minor-merger event. {\color{black}An analysis of the angular momentum budget indicates that the angular momentum lost by the stars during these mergers is largely transferred to the dark matter halo.}

The rapid formation of these spheroids is in contrast to the average morphological change of the general massive spheroid population \citep[Figure 11 in][]{Martin2018b}, in which the mean $\sfrac{V}{\sigma}$ changes from the disc regime to values close to 0.3 over Hubble time (note, however, that this is not the same as the average change in $\sfrac{V}{\sigma}$ in individual typical spheroids where the morphological change can be bursty and changes fastest during merger events). For the spheroids in this study, a similar change in $\sfrac{V}{\sigma}$ is achieved within only a few hundred Myrs.

\begin{center}
\begin{table}
\centering
\begin{tabular}{|| c | c | c | c ||}
1 & 2 & 3 & 4\\
\hline
ID & M$_*$ & Rel. f$_{gas}$ & Rel. clumpiness \\
& (10$^{11}$M$_\odot$) & &\\
\hline
\hline
1841    & 1.16  & 1.15$\pm0.13$  & 0.79$\pm{0.05}$ \\
5119    & 0.94  & 0.99$\pm0.15$  & 1.17$\pm{0.08}$ \\
1078    & 0.21  & 0.89$\pm0.36$  & 0.84$\pm{0.06}$ \\
1956    & 0.40  & 1.19$\pm0.27$  & 1.55$\pm{0.11}$ \\
8233    & 0.72  & 1.23$\pm0.17$  & 1.61$\pm{0.11}$ \\
9675    & 0.57  & 1.53$\pm0.26$  & 1.65$\pm{0.12}$ \\
18163   & 0.63  & 1.20$\pm0.16$  & 2.06$\pm{0.15}$ \\
3341    & 0.23  & 1.08$\pm0.39$  & 1.22$\pm{0.23}$ \\
184     & 1.01  & 1.42$\pm0.20$  & 0.74$\pm{0.05}$ \\
952     & 0.42  & 1.47$\pm0.31$  & 0.62$\pm{0.04}$
\end{tabular}
\caption{Properties of the more massive progenitor of the spheroids studied in this paper. Columns: (1) Galaxy ID, (2) Stellar mass in units of 10$^{11}$M$_\odot$, (3), (4) Gas fraction and stellar clumpiness respectively, relative to the mean of a control sample at the merger redshift.}
\label{progenitor properties}
\end{table}
\end{center}

\begin{figure}
\includegraphics[width=0.45\textwidth]{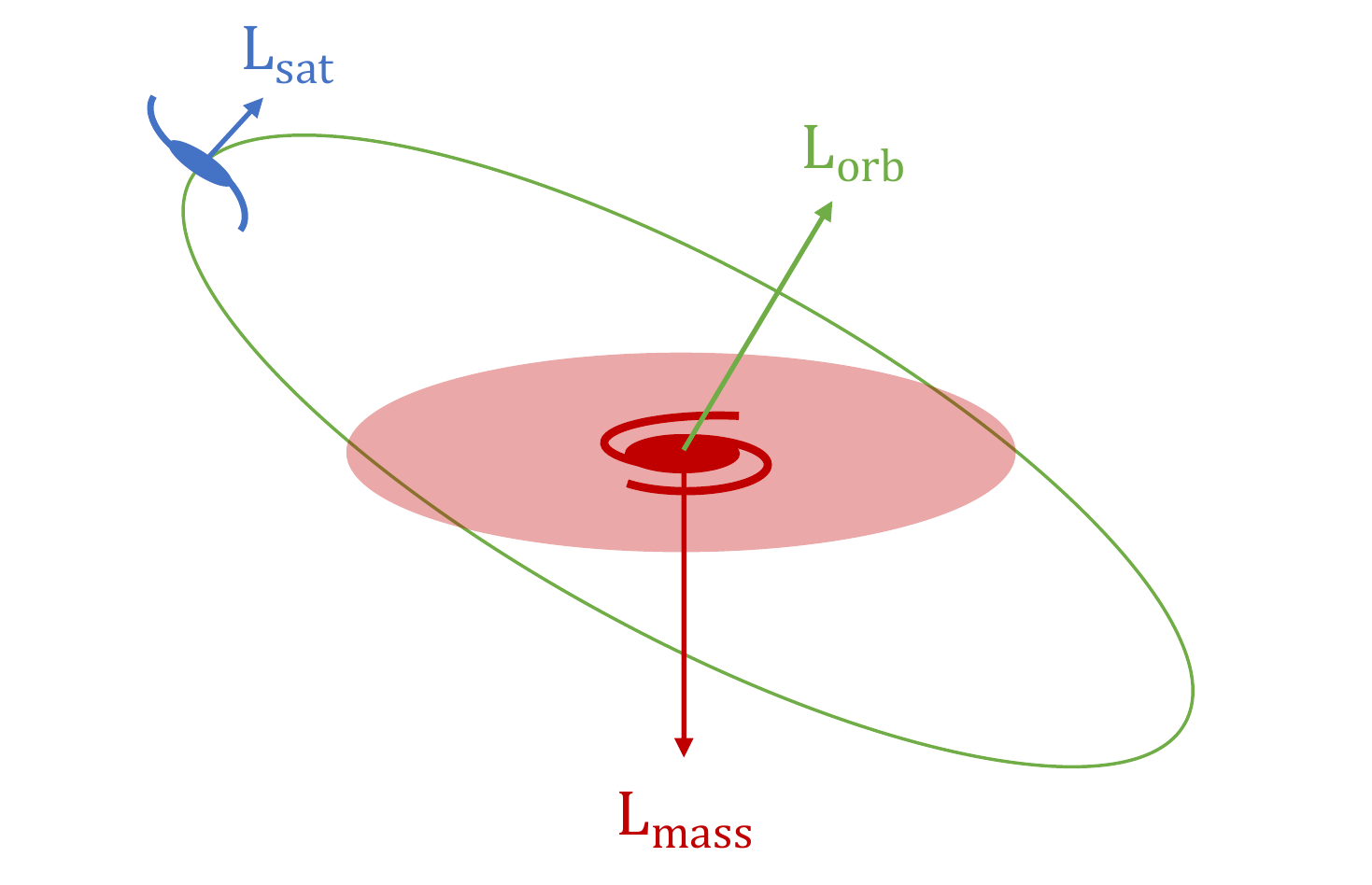}
\caption{Diagram indicating the vectors used to describe the orbital configuration of the mergers. The angles between vectors are defined in the standard way, in the common plane containing both vectors.}
\label{gal_diagram}
\end{figure}

The middle panels show the change in the stellar mass of the galaxy (solid line), the ex-situ i.e. accreted stellar mass (dashed line) and the gas mass of the system (dotted line). The step change in the ex-situ mass (dashed line) around the time of the merger indicates the stellar mass brought in by the accreted satellite. The galaxies undergo strong stellar mass growth during the merger, with high star-formation rates (demonstrated by the steep gradient of the solid line), a corresponding decrease in the gas mass as it is consumed as the merger progresses, and an increase in the effective radius (right-hand panels). In many cases the gas mass increases prior to the star formation event, due to gas brought into the system by the satellite. It is worth noting that these merger remnants do not undergo compaction, akin to minor-merger-driven disc instabilities that are expected to take place in the gas-rich high redshift Universe \citep{Dekel14,Zolotov2015,vanDokkum2015} Instead, they continue to grow in stellar mass and size after the merger. 

\section{Properties of the progenitor system}
\label{progenitor}

We proceed by exploring why these minor mergers have such a profound and unusual impact on the morphology of the remnant, by studying properties of the progenitor system that are likely to contribute to the observed evolution of the morphology, stellar mass and effective radius of the remnant. We focus on three characteristics: the gas fraction and clumpiness of the more massive progenitor and the orbital configuration of the merging galaxies. 

Disks in the early Universe ($z>2$) can be gas-rich and clumpy and more susceptible to gravitational instabilities that can be triggered by events such as minor mergers \citep[e.g.][]{Swinbank2012, Zolotov2015, Saha2018}. Even though the redshifts of the mergers that create our spheroids are typically much lower than the epoch of such gas-rich disks, it is worth considering whether the more massive progenitors may be anomalously gas-rich and/or clumpy, compared to control samples at similar redshifts. Note that we have also examined whether the merging satellites have properties that diverge from other galaxies of similar stellar masses at the merger redshifts in question, but find that they are not anomalous. In any case, given the typically low mass ratios of these events, the properties of the spheroidal remnants are driven more by those of the more massive progenitors, which are, therefore, more pertinent to our analysis.  

In a similar vein, the outcome of a merger can be dependent on the orbital configuration of the progenitor system \citep[e.g.][]{Naab2003,Chilingarian2010,Solanes2015,Martin2018b}. It is, therefore, worth exploring whether the progenitors of these spheroids show a pattern in their relative trajectories that may explain the properties of their remnants. 

\subsection{Gas fraction and clumpiness of the more massive progenitor}
\label{gas_fraction}

We begin by comparing the gas fraction and clumpiness of the more massive progenitor, at the timestep before the merger, to a control sample of 200 randomly selected disc galaxies at the same redshift which have stellar masses within 10 per cent of the galaxy in question. The gas fraction is defined as:

\begin{equation}
f_{gas} = \frac{M_{gas}}{M_{\star} + M_{gas}},
\end{equation}

where $M_{\rm{gas}}$ and $M_{\star}$ are the total gas and stellar masses in the progenitor system respectively. Column 3 in Table \ref{progenitor properties} shows the ratio of the gas fractions in the more massive progenitors just before the merger to those in the control samples. For many systems these ratios are actually lower than the median values for similar disc galaxies at the same redshift, while for the others the ratios are less than $\sim$1.5. Compared to galaxies at intermediate redshift (where most of our spheroids form), gas fractions in the gas-rich early Universe are expected to be almost an order of magnitude larger \citep[see Figure 4 in][]{Martin2018b}, indicating that, on the whole, the more massive progenitors in these mergers are not anomalously gas-rich. 

We proceed by comparing the clumpiness of the more massive progenitor to the control samples\footnote{Note that, given the spatial resolution of the simulation, we are largely exploring kpc-scale clumps \citep{Elmegreen2013,Cava2018} in this exercise.}. {\color{black}Since we are interested in the \textit{relative} clumpiness, compared to that of other galaxies at the same redshift, we simply use the star-particle distributions for this exercise.} To measure the clumpiness in the stars we follow \citet{2003ApJS..147....1C} and \citet{2011MNRAS.418..801H}. {\color{black}We first apply a Gaussian filter with a standard deviation of $0.3\times r_{20}$, where r$_{20}$ is the radius that contains 20 per cent of the galaxy's stellar mass surface density \citep{2003ApJS..147....1C}, and then subtract the original image from the smoothed one, which leaves the high-frequency structure as a residual. The absolute signal in this residual image is then summed and divided by the sum of that in the original image to yield a value for the clumpiness.}

We study the ratio of the clumpiness of the more massive progenitors to control samples at the redshift of the merger. Column 4 in Table \ref{progenitor properties} shows that many of the progenitor systems are  substantially \textit{less} clumpy than their control samples and that the clumpiness of the more massive progenitors is within 46 per cent of the mean values of the control samples in all cases. On the whole, the ratios do not indicate that the more massive progenitors of our spheroids have anomalously high values of clumpiness. 

Hence, the progenitor systems do not appear to be atypical, either in terms of their gas fractions or their clumpiness. These properties are, therefore, unlikely to be the key factors behind the rapid morphological transformation observed in these mergers.

\subsection{Orbital configurations}
\label{orbits}
We proceed by investigating whether the merger progenitors may have a specific orbital configuration which plays a role in the observed properties of the remnant. The orbital configurations are calculated at the closest timestep before coalescence. 
Figure \ref{gal_diagram} describes the vectors used to quantify the orbital configuration of the mergers. We study two key aspects of the orbital configuration: the angle between the spin (i.e. angular momentum) vectors of the two merging galaxies (i.e. whether the merger is prograde or retrograde) and the angle between the angular momentum vector of the satellite orbit and the spin of the more massive galaxy.

\begin{figure}
\includegraphics[width=\columnwidth]{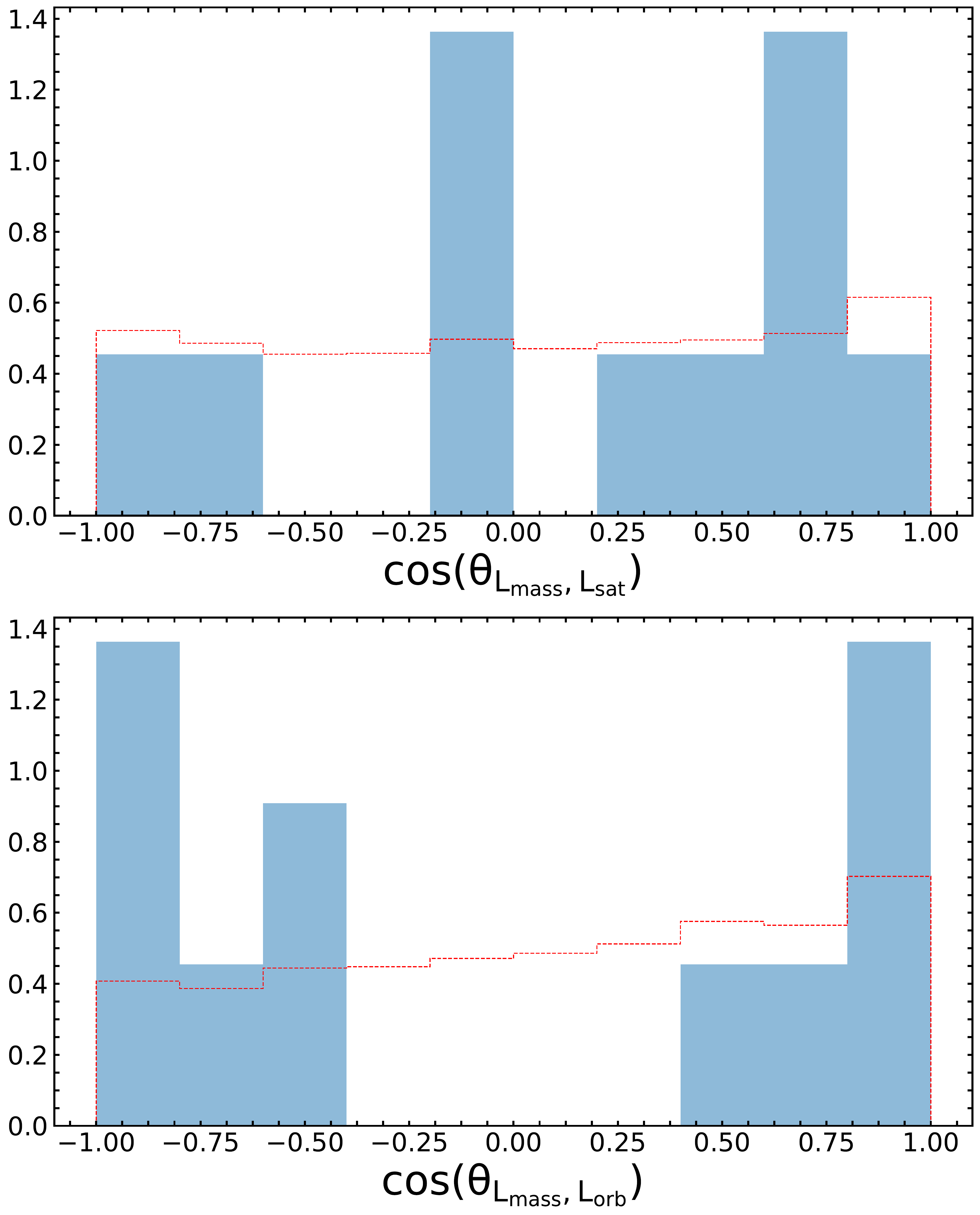}
\caption{Top: The cosine of the angle between the spin vectors, defined using the star particles, of the massive galaxy ($L_{mass}$) and the satellite ($L_{sat}$). The corresponding distribution from a control sample is shown in red (the normalisation is arbitrary and chosen for clarity). There is no preference for the spins of the two galaxies to be either prograde or retrograde. Bottom: The cosine of the angle between $L_{orb}$ (see Figure \ref{gal_diagram}) and the spin vector of the massive galaxy ($L_{mass}$). The merger events cluster around a specific orbital configuration, where the satellite's orbit tends to be close to co-planar with the disk of the massive galaxy i.e. $\cos{\theta}$ tends to be close to either -1 or +1.}
\label{vectors}
\end{figure}

The top panel of Figure \ref{vectors} shows the alignment of the spin vectors, defined using the star particles, of the two galaxies. Positive values of $\cos (\theta)$ indicate prograde mergers, where the spin vectors of the discs of the two galaxies point in the same direction. Negative values of $\cos (\theta)$ indicate retrograde mergers, where the spin vectors of the discs of the two galaxies do not point in the same direction. The red dotted line indicates the distribution from a control sample (constructed as described in Section \ref{gas_fraction} above). As might be expected, neither the control sample nor the progenitors of our spheroids show any preference for prograde or retrograde mergers. The relative spins of the two merging galaxies are therefore not the principal cause of the properties of the remnant.

Finally, we study the orientation of the orbit of the satellite relative to the disc of the more massive galaxy. We first measure the angular momentum vector of the satellite orbit, defined as $L_{orbit} = M_{sat}.(r\times v)$, where $v$ and $r$ are the velocity and position vectors of the satellite with respect to the massive galaxy respectively, and then calculate the angle between $L_{orbit}$ and the spin vector of the massive progenitor. Thus, a value of 0 for $\cos{\theta}$ represents a satellite whose orbit is exactly perpendicular to the disc of the more massive progenitor. A value of 1 indicates that the satellite orbit is exactly co-planar with the plane of the disc of the more massive progenitor and prograde with its spin, while a value of -1 indicates that the satellite orbit is co-planar with the plane of the disc of the more massive progenitor, but retrograde with its spin. 

We find that, unlike the control sample which exhibits a flat distribution in orbital configurations (although it shows a slight, expected tendency towards prograde co-planar orbits, as satellites in the inner parts of halos tend to align with the central galactic plane, see \citet{Welker2018}), the progenitors of our spheroids strongly favour satellite trajectories that are close to co-planar with the disc of the more massive progenitor (regardless of whether the mergers are prograde or retrograde). Such trajectories are likely to be those that are able to deliver the orbital energy of the satellite into the disc of the more massive progenitor in the most efficient manner \citep[e.g.][]{Cox2008}, which likely enables even a small satellite (recall that the lowest mass ratio in the progenitors of our spheroids is 0.11) to quickly destroy the disc of the massive progenitor and create a dispersion-dominated system. 

\begin{figure}
\includegraphics[width=\columnwidth]{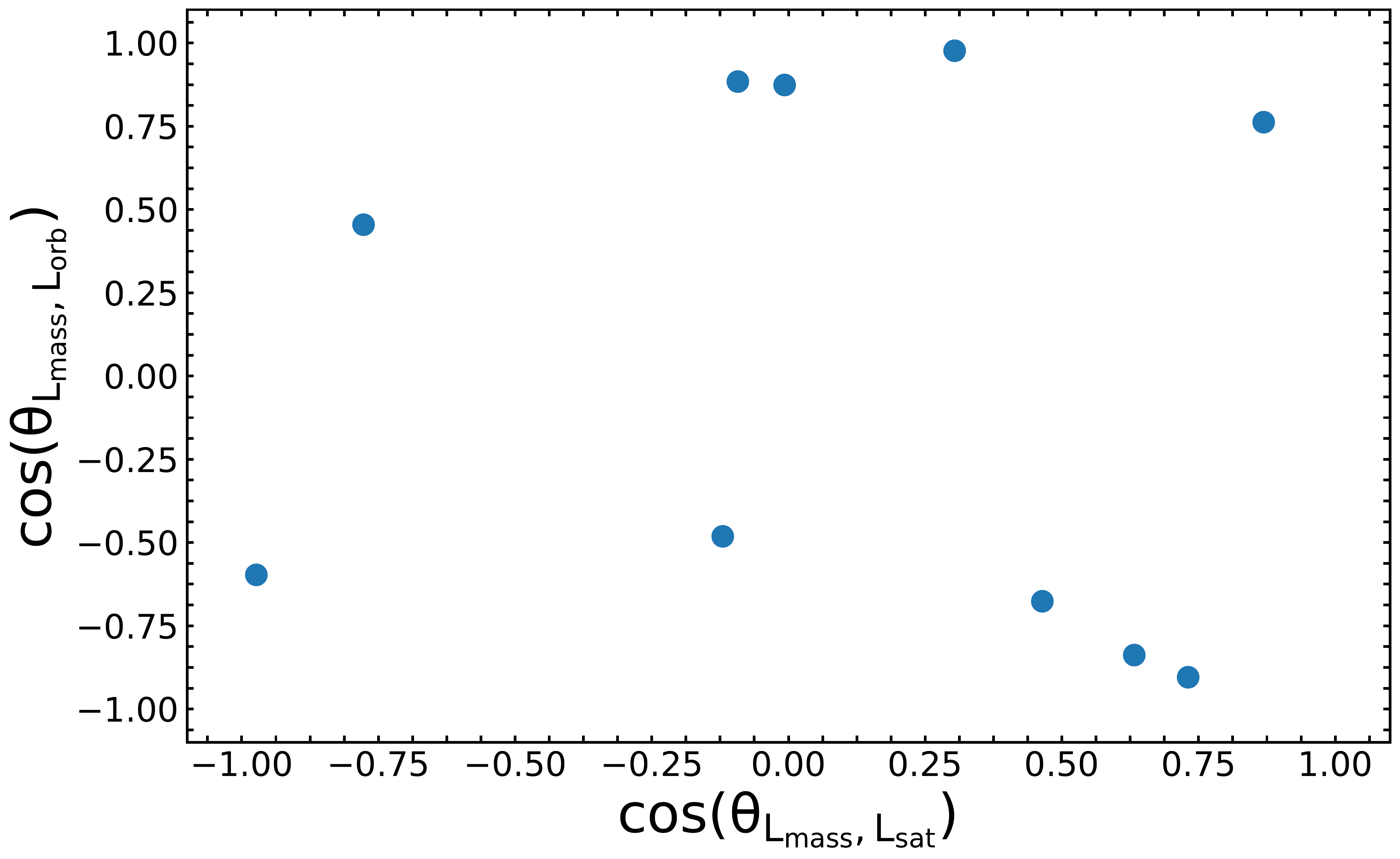}
\caption{The cosine of the angle between the (stellar) spin vectors of the massive galaxy ($L_{mass}$) and the satellite ($L_{sat}$) plotted against the cosine of the angle between $L_{orb}$ (see Figure \ref{gal_diagram}) and the spin vector of the massive galaxy ($L_{mass}$). }
\label{spinvsorbit}
\end{figure}

{\color{black}In Figure \ref{spinvsorbit}, we combine the information in the two histograms from Figure \ref{vectors} by plotting the angles between the spin vectors of the two galaxies against the angle between the satellite orbit and the spin of the massive galaxy. Our systems show no preference for either the relative spins of the two galaxies or the satellite orbit to be prograde or retrograde. In addition, there appears to be no correlation between these two quantities. These mergers therefore occur in all combinations of spin and orbit alignment, with the only common feature being that they are co-planar.}

Since our galaxy sample is relatively small, we briefly check whether a broad similar trend remains if we relax the mass threshold of the analysis. As we show in the Appendix, studying the orbital configurations of similar (i.e. low $f_{exsitu}$) spheroids with a mass cut of $10^{10.5}$M$_\odot$ shows the same general tendency towards co-planar orbits, indicating that the result for our M$_*$$>$10$^{11}$M$_\odot$ does not occur by chance. Note, however, that we perform this exercise only as a sanity check on our results; lower mass galaxies can naturally end up with a large in-situ stellar mass fraction via early gas accretion and not due to a peculiarity in their merger histories \citep{Martin2018b}, as a result of which the deficit of non-planar orbits in Figure \ref{mass10_5} is not as severe for lower masses, even though they have low $f_{exsitu}$. Our focus therefore remains on the most massive galaxies (M$_*$$>$10$^{11}$M$_\odot$), beyond the knee of the mass function, where low $f_{exsitu}$ is particularly anomalous. 

{\color{black} Finally, it is also interesting to consider whether a disc could survive a co-planar orbital configuration similar to the ones that produce our population of spheroids. We find that there are some discs which have similar stellar masses as the spheroids in our study which undergo coplanar mergers but maintain their discy morphology. These systems overwhelmingly prefer \emph{prograde} minor mergers. In addition, the mergers have higher gas fractions ($\times$1.7 the control sample compared to $\times$1.1 for the spheroids) and lower mass ratios (median mass ratio of 0.12 compared to 0.2 for the spheroids). The survival of the disc is therefore due to a combination of a low mass ratio merger which is relatively gas-rich and has a prograde configuration. This disc population will be studied in detail in a forthcoming paper (Jackson et al. in prep).}




\section{Summary} 
\label{conc}
The most massive galaxies ($M_*>10^{11}$M$_\odot$), regardless of morphology, typically exhibit rich merger histories, which results in high fractions of `ex-situ' stellar mass i.e. mass accreted directly from external objects rather than having been formed in-situ. This is particularly true for the most massive spheroids at the present day, whose stellar masses have classically been considered unattainable without significant numbers of both major and minor mergers. 

In this paper, we have studied the formation of a sample of massive ($M_*>10^{11}$M$_\odot$) spheroidal systems, in the Horizon-AGN simulation, which have anomalously low ($<0.3$) ex-situ mass fractions. These systems are unusual, in the sense that they are extremely massive and spheroidal, yet do not exhibit significant fractions of ex-situ stellar mass - the opposite to what is usually the case for such galaxies.  

We have shown that these objects are created in a single minor merger, which triggers a catastrophic (and permanent) morphological change, with $\sfrac{V}{\sigma}$ declining rapidly from the disc to spheroid regime in only a few hundred Myrs. The merger event triggers strong star formation, driven by the native gas reservoir of the more massive progenitor, and an increase in the effective radius of the system. The remnant is a massive, relatively diffuse, slowly-rotating spheroidal galaxy. 

We have studied the properties of the progenitor system to explore why the minor merger has such a profound effect on the remnant. Since minor mergers are thought to be able to influence the production of spheroidal systems by triggering gravitational instabilities in clumpy, gas-rich disc galaxies at high redshift, and since orbital configurations can have an impact on the properties of the remnant, we have focussed on the evolution of these properties in the progenitor systems of our spheroids.

We have shown that the properties of the individual progenitors are not anomalous. The more massive progenitors in these mergers have gas fractions and clumpiness values that are consistent with the mean values in control samples at the same redshifts. However, the relative trajectories of the merging progenitors exhibit a specific orbital configuration that likely drives the observed behaviour. The satellites approach the more massive galaxies in close to co-planar orbits, which are likely to maximize the tidal forces and therefore the transfer of orbital energy, enabling the satellite to destroy the disc in only a few hundred Myrs and drive the creation of a massive spheroidal galaxy.  

{\color{black} The principal distinguishing feature of the spheroids described in this paper is the high SFR of these systems for a time after the spheroid has formed. Recent ultraviolet studies have led to the the discovery of widespread star formation activity in spheroidal galaxies across cosmic time \citep[see e.g.][]{Yi2005,Kaviraj2007}. However, this star formation is typically at a low level \citep[e.g.][]{Kaviraj2008}, with spheroidal galaxies lying well below the star-formation main sequence \citep[e.g.][]{Salim2007,Kaviraj2013}, even if they have had recent mergers. This is particularly true for spheroids in the low and intermediate redshift Universe.} 

Our spheroids on the other hand lie significantly above the star-formation main sequence (column 11 in Table \ref{spheroid properties}) which offers a route to potentially identifying such systems in observations. Indeed, some (rare) cases of spheroids with SFRs well above the star formation main sequence are known \citep[e.g.][]{fukugita04,schawinski10}, with number fractions of a few per cent (similar to our findings here). {\color{black}While such spheroids are difficult to explain via traditional channels, it is plausible that at least some of these massive spheroids with anomalously high SFRs are formed via the co-planar minor-merger channel described here.}


Our study adds to the burgeoning literature that is highlighting the significant role of processes other than major mergers in driving the change in the morphological mix of massive galaxies. It demonstrates that, for close to co-planar orbital configurations, a single minor merger (with a mass ratio as low as $\sim$1:10) is able to completely destroy a disc and create a spheroid, even at the high-mass end of the galaxy mass function. While this channel produces a minority ($\sim$5 per cent) of such galaxies, our study indicates that the formation of the \textit{most} massive spheroids in the local Universe need not involve any major mergers -- or indeed any significant merging at all -- contrary to our classical assumptions. 

\section*{Acknowledgements}
We are grateful to the referee, Dan Taranu, for numerous suggestions that significantly improved the original manuscript. RAJ acknowledges support from the STFC [ST/R504786/1]. GM acknowledges support from the STFC [ST/N504105/1]. SK acknowledges a Senior Research Fellowship from Worcester College Oxford. CL is supported by a Beecroft Fellowship. JD acknowledges funding support from Adrian Beecroft, the Oxford Martin School and the STFC. This research has used the DiRAC facility, jointly funded by the STFC and the Large Facilities Capital Fund of BIS, and has been partially supported by grant Spin(e) ANR-13-BS05-0005 of the French ANR. This work was granted access to the HPC resources of CINES under the allocations 2013047012, 2014047012 and 2015047012 made by GENCI. This work is part of the Horizon-UK project.




\bibliographystyle{mnras}
\bibliography{bib} 



\appendix

\section{Orbital configurations with a relaxed mass cut}
Section \ref{orbits} indicates that the principal difference in the formation mechanism of our massive, low $f_{exsitu}$ spheroids compared to typical (i.e. high $f_{exsitu}$) spheroids of similar mass is that they form via minor mergers where the satellite orbit is close to being co-planar with the disc of the massive galaxy. As noted in Section \ref{orbits}, since our sample size is relatively small, we check whether this result could have originated by chance, by exploring whether this tendency towards co-planar orbits persists with more relaxed selection criteria. 

\begin{figure}
\includegraphics[width=\columnwidth]{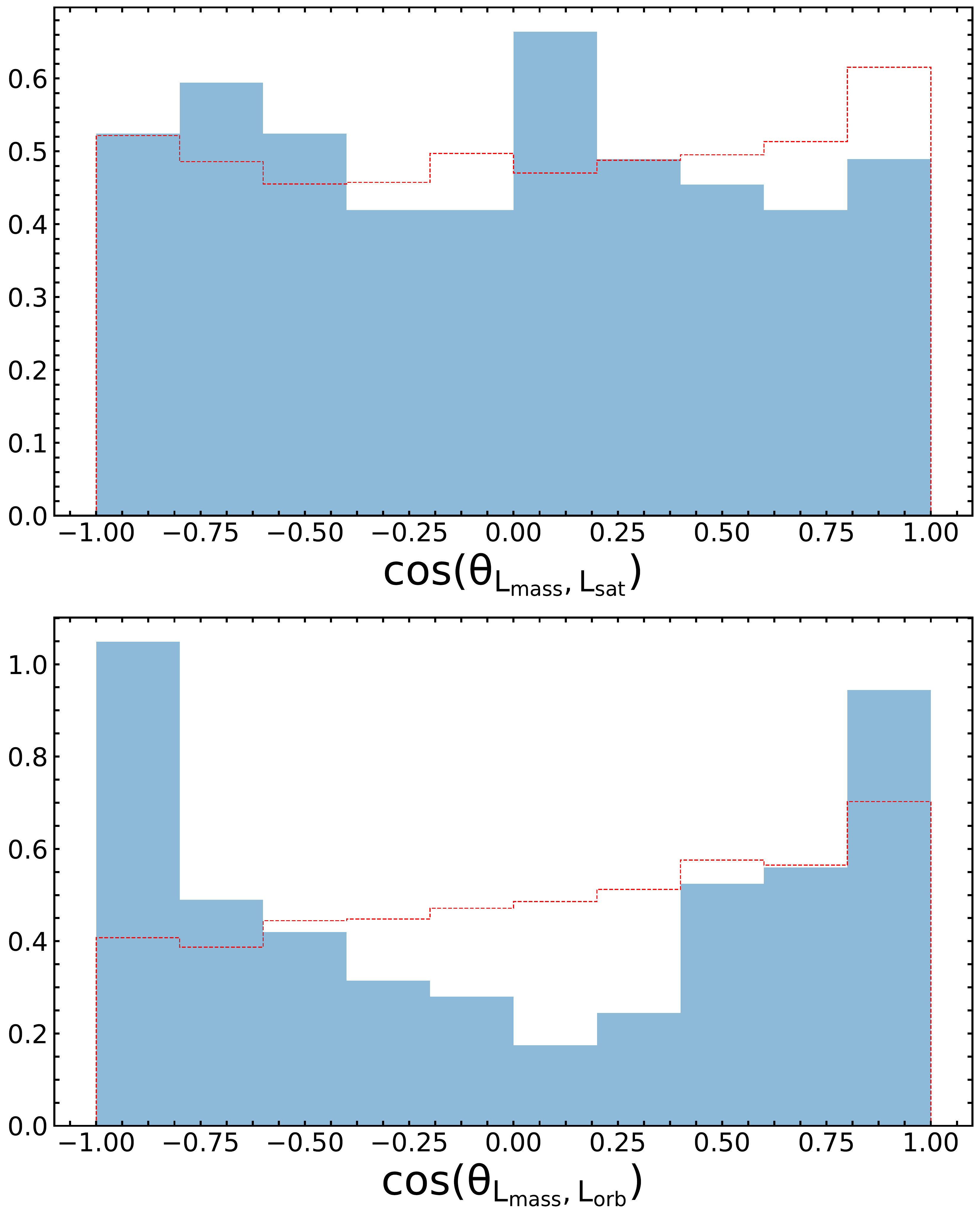}
\caption{Same as Figure \ref{vectors} but for a sample where the mass threshold is M$_*$ $>$ 10$^{10.5}$ M$_{\odot}$, with other parameters unchanged (i.e. $\sfrac{V}{\sigma}<0.3$ and $f_{exsitu}<0.3$). As is the case for the M$_*$ $>$ 10$^{11}$ M$_{\odot}$, low $f_{ex-situ}$ spheroids continue to show a tendency towards mergers that have close to co-planar orbits.}
\label{mass10_5}
\end{figure}

For this exercise, we reduce the mass threshold of our sample to M$_*$$>$10$^{10.5}$M$_\odot$ but leave constraints on the other variables (i.e. $\sfrac{V}{\sigma}<0.3$ and $f_{exsitu}<0.3$) the same. We perform an identical orbital configuration analysis on this sample as that described in Section \ref{orbits}. As Figure \ref{mass10_5} indicates this analysis shows the same tendency towards close-to-co-planar orbits creating low $f_{exsitu}$ spheroids, indicating that the result for the more massive spheroids does not occur by chance.

Note, however, that, as described in Section \ref{orbits}, we use this only as a sanity check of our results. Lower-mass galaxies naturally produce larger in-situ mass fractions because they produce more mass via gas accretion, as opposed to mergers \citep{Martin2018b}. As a result, the deficit of non co-planar orbits in Figure \ref{mass10_5} is not as severe in this lower-mass regime as it is for their more massive counterparts which are the subject of this paper. Our focus, therefore, remains on the most massive galaxies (M$_*$$>$10$^{11}$M$_\odot$) in which low $f_{exsitu}$ is particularly anomalous.

\bsp	
\label{lastpage}
\end{document}